\documentclass{article}
\usepackage{amsmath}
\usepackage{amssymb}
\usepackage{bm}

\title{\Large\bf Notes on the Dipole Coordinate System}
\author{M. Swisdak\footnote{swisdak@ppd.nrl.navy.mil} \\
\it \normalsize{Plasma Physics Division, Naval Research Laboratory,
Washington, DC, 20375} }
\date{}

\begin{document}
\maketitle

\begin{abstract}
A strong magnetic field can make it advantageous to work in a
coordinate system aligned with dipolar field lines.  This monograph
collect the formulas for some of the most frequently used expressions
and operations in dipole coordinates.
\end{abstract}

\bigskip

In some physical systems ({\it e.g.}, the terrestrial ionosphere or
the solar corona) the magnetic field can impose a strong anisotropy by
restricting transport processes perpendicular to the lines of force.
If the field is approximately dipolar it can be useful to work in an
aligned coordinate system even though the vector operations are
somewhat more complicated than in a Cartesian or spherical polar
representation.  Here we present several of the formulas that
frequently occur when working with vectors in dipolar coordinates.
Although some of the results have previously appeared in the
literature \cite{fatkullin72a,orens79a}, others seem to be new.

There are many possible (related) choices for the dipolar
coordinates.  The right-handed orthogonal system considered here,
$(q,p,\phi)$, is defined in terms of the usual spherical polar
coordinates, $(r,\theta,\phi)$, by
\begin{equation}\label{def}
q = \frac{\cos{\theta}}{r^2} \qquad
p = \frac{r}{\sin^2\theta} \qquad \phi = \phi
\end{equation}
$p$ is constant along a dipolar field line while $q$ parameterizes the
displacement parallel to the field: $q=0$ at the equator, $q
\rightarrow -\infty$ as $\theta \rightarrow \pi$ and $q \rightarrow
+\infty$ as $\theta \rightarrow 0$.

\section{Inverse Transformation}

The inversion of (\ref{def}) --- {\it i.e.}, finding
$(r,\theta)$ given $(q,p)$ --- involves the solution of a non-trivial
equation.  Substituting for $r$ gives $qp^2 =
\cos\theta/\sin^4\theta$ while eliminating $\theta$ leads to the
polynomial expression
\begin{equation}\label{poly}
q^2r^4 + \frac{r}{p}-1=0
\end{equation}
Descartes's rule of signs states that (\ref{poly}) has exactly one
positive, real root and, as has been previously noted
\cite{kageyama06a}, since (\ref{poly}) is a quartic this root his an
algebraic representation.  To be useful in numerical models however
such a solution has to be expressed in a computationally stable form.
Define the auxiliary quantities
\begin{equation}
  \alpha = \frac{256}{27}q^2p^4 \qquad \beta = ( 1 + \sqrt{1 + \alpha}
  )^{2/3} \qquad \gamma = \sqrt[3]{\alpha}
\end{equation}
and
\begin{equation}
\mu = \frac{1}{2}\left(\frac{\beta^2 + \beta\gamma +
  \gamma^2}{\beta}\right)^{3/2}
\end{equation}
Then the positive real root of (\ref{poly}) is
\begin{equation}\label{sol}
  r = \frac{4\mu}{(1+\mu)( 1+\sqrt{2\mu-1})}\,p
\end{equation}
Since $\mu \geq 1$ this formulation eliminates the possibility of a
catastrophic cancellation between terms. To complete the inversion
recall that $\sin^2\theta = r/p$.

Although (\ref{sol}) makes it possible, in principle, to
express formulas in terms of either $r$ and $\theta$ or $q$ and $p$,
it is usually simpler to use the former representation.

\section{Coordinate Derivatives}

In this section we ignore the $\phi$ coordinate.  The partial
derivatives of the dipolar coordinates with respect to the spherical
polar coordinates are
\begin{alignat}{2}
\frac{\partial q}{\partial r} &= -\frac{2\cos\theta}{r^3} & \qquad
\frac{\partial p}{\partial r} =& \frac{1}{\sin^2\theta} \\
\frac{\partial q}{\partial \theta} &= -\frac{\sin\theta}{r^2} & \qquad
\frac{\partial p}{\partial \theta} =& -\frac{2r\cos\theta}{\sin^3\theta}
\end{alignat}
The Jacobian is then
\begin{equation}\label{jac}
\frac{\partial (q,p)}{\partial (r,\theta)} = \frac{\delta^2}{r^2\sin^3\theta}
\end{equation}
where $\delta(\theta) = \sqrt{1+3\cos^2\theta}$.  The derivatives of
the spherical polar coordinates with respect to the dipolar
coordinates are
\begin{alignat}{2}
\frac{\partial r}{\partial q} &= -\frac{2r^3\cos\theta}{\delta^2} &
\qquad \frac{\partial r}{\partial p} &= \frac{\sin^4\theta}{\delta^2}
\\
\frac{\partial \theta}{\partial q} &= -\frac{r^2\sin\theta}{\delta^2} &
\qquad \frac{\partial \theta}{\partial p} &=
-\frac{2\cos\theta\sin^3\theta}{r\delta^2}
\end{alignat}
A few second derivatives are also occasionally useful
\begin{alignat}{2}
\frac{\partial^2 r}{\partial q^2} &=
-\frac{2r^5}{\delta^6}(1-10\cos^2\theta-15\cos^4\theta) &\qquad
\frac{\partial^2 r}{\partial p^2} &=
-\frac{4\sin^6\theta\cos^2\theta}{r\delta^4}(5+3\cos^2\theta) \\
\frac{\partial^2\theta}{\partial q^2} &=
\frac{r^4\sin\theta\cos\theta}{\delta^6}(11+9\cos^2\theta) &\qquad
\frac{\partial^2\theta}{\partial p^2} &=
-\frac{2\sin^5\theta\cos\theta}{r^2\delta^6}(1-16\cos^2\theta-9\cos^4\theta)
\end{alignat}

\section{Unit Vectors}

The expressions for the dipolar and spherical polar unit vectors in
the alternate coordinate system are
\begin{alignat}{3}
\mathbf{\hat{q}} &= -\frac{2\cos\theta}{\delta}\mathbf{\hat{r}} -
\frac{\sin\theta}{\delta}\mathbf{\hat{\boldsymbol{\theta}}} & \qquad
\mathbf{\hat{p}} &=& &\frac{\sin\theta}{\delta}\mathbf{\hat{r}} -
\frac{2\cos\theta}{\delta}\mathbf{\hat{\boldsymbol{\theta}}} \\ 
\mathbf{\hat{r}} &= -\frac{2\cos\theta}{\delta}\mathbf{\hat{q}} +
\frac{\sin\theta}{\delta}\mathbf{\hat{p}} & \qquad
\mathbf{\hat{\boldsymbol{\theta}}} &=&
-&\frac{\sin\theta}{\delta}\mathbf{\hat{q}} -
\frac{2\cos\theta}{\delta}\mathbf{\hat{p}}
\end{alignat}

\subsection{First Derivatives of Unit Vectors}
\begin{alignat}{6}
\frac{\partial \mathbf{\hat{r}}}{\partial q} &=&
&\mathbf{\hat{\boldsymbol{\theta}}}\frac{\partial \theta}{\partial q}
&\qquad \frac{\partial \mathbf{\hat{r}}}{\partial p} &=&
&\mathbf{\hat{\boldsymbol{\theta}}}\frac{\partial \theta}{\partial p}
&\qquad \frac{\partial \mathbf{\hat{r}}}{\partial \phi} &=&
&\mathbf{\hat{\boldsymbol{\phi}}}\sin\theta \\
\frac{\partial \mathbf{\hat{\boldsymbol{\theta}}}}{\partial q} &=&
-&\mathbf{\hat{r}}\frac{\partial \theta}{\partial q} &\qquad
\frac{\partial \mathbf{\hat{\boldsymbol{\theta}}}}{\partial p} &=&
-&\mathbf{\hat{r}}\frac{\partial \theta}{\partial p} &\qquad
\frac{\partial \mathbf{\hat{\boldsymbol{\theta}}}}{\partial \phi} &=&
&\mathbf{\hat{\boldsymbol{\phi}}}\cos\theta \\
\frac{\partial \mathbf{\hat{\boldsymbol{\phi}}}}{\partial q} &=& &0
&\qquad \frac{\partial \mathbf{\hat{\boldsymbol{\phi}}}}{\partial p}
&=& &0 &\qquad \frac{\partial
\mathbf{\hat{\boldsymbol{\phi}}}}{\partial \phi} &=&
-&\mathbf{\hat{r}}\sin\theta
-\mathbf{\hat{\boldsymbol{\theta}}}\cos\theta
\end{alignat}
and
\begin{alignat}{6}
\frac{\partial \mathbf{\hat{q}}}{\partial q} &=&
&\mathbf{\hat{p}}\frac{3(1+\cos^2\theta)}{\delta^2}\frac{\partial
\theta}{\partial q} &\qquad
\frac{\partial \mathbf{\hat{q}}}{\partial p} &=&
&\mathbf{\hat{p}}\frac{3(1+\cos^2\theta)}{\delta^2}\frac{\partial
\theta}{\partial p} &\qquad
\frac{\partial \mathbf{\hat{q}}}{\partial \phi} &=&
-&\mathbf{\hat{\boldsymbol{\phi}}}\frac{3\cos\theta\sin\theta}{\delta}
\\
\frac{\partial \mathbf{\hat{p}}}{\partial q} &=&
-&\mathbf{\hat{q}}\frac{3(1+\cos^2\theta)}{\delta^2}\frac{\partial
\theta}{\partial q} &\qquad
\frac{\partial \mathbf{\hat{p}}}{\partial p} &=&
-&\mathbf{\hat{q}}\frac{3(1+\cos^2\theta)}{\delta^2}\frac{\partial
\theta}{\partial p} &\qquad
\frac{\partial \mathbf{\hat{p}}}{\partial \phi} &=&
&\mathbf{\hat{\boldsymbol{\phi}}}\frac{1-3\cos^2\theta}{\delta} \\
\frac{\partial \mathbf{\hat{\boldsymbol{\phi}}}}{\partial q} &=& &0
&\qquad \frac{\partial \mathbf{\hat{\boldsymbol{\phi}}}}{\partial p}
&=& &0 &\qquad \frac{\partial
\mathbf{\hat{\boldsymbol{\phi}}}}{\partial
\phi} &=& &\mathbf{\hat{q}}\frac{3\cos\theta\sin\theta}{\delta}
-\mathbf{\hat{p}}\frac{1-3\cos^2\theta}{\delta}
\end{alignat}

\subsection{Some Second Derivatives}
\begin{alignat}{2}
\frac{\partial^2\mathbf{\hat{q}}}{\partial q^2} =
-&\mathbf{\hat{q}}\left(\frac{3(1+\cos^2\theta)}{\delta^2}\frac{\partial
\theta}{\partial q}\right)^2 -
&&\mathbf{\hat{p}}\frac{3r^2\cos\theta}{\delta^6}
(15+16\cos^2\theta+9\cos^4\theta)
\frac{\partial \theta}{\partial q}\\
\frac{\partial^2\mathbf{\hat{q}}}{\partial p^2} =
-&\mathbf{\hat{q}}\left(\frac{3(1+\cos^2\theta)}{\delta^2}\frac{\partial
\theta}{\partial p}\right)^2 +
&&\mathbf{\hat{p}}\frac{3\sin^2\theta}{r\delta^6}(1-23\cos^2\theta
-17\cos^4\theta-9\cos^6\theta)\frac{\partial
\theta}{\partial p} \\
\frac{\partial^2\mathbf{\hat{p}}}{\partial q^2} =
-&\mathbf{\hat{p}}\left(\frac{3(1+\cos^2\theta)}{\delta^2}\frac{\partial
\theta}{\partial q}\right)^2 +
&&\mathbf{\hat{q}}\frac{3r^2\cos\theta}{\delta^6}(15+16\cos^2\theta
+9\cos^4\theta)\frac{\partial \theta}{\partial q} \\
\frac{\partial^2\mathbf{\hat{p}}}{\partial p^2} =
-&\mathbf{\hat{p}}\left(\frac{3(1+\cos^2\theta)}{\delta^2}\frac{\partial
\theta}{\partial p}\right)^2 -
&&\mathbf{\hat{q}}\frac{3\sin^2\theta}{r\delta^6}(1-23\cos^2\theta
-17\cos^4\theta-9\cos^6\theta)\frac{\partial \theta}{\partial p}
\end{alignat}

\begin{equation}
\frac{\partial^2\mathbf{\hat{\boldsymbol{\phi}}}}{\partial \phi^2} =
-\mathbf{\hat{\boldsymbol{\phi}}}
\end{equation}

\section{Metric; Differential Line, Area, and Volume Elements}
Since the dipole system is orthogonal the only non-zero components of
the metric are the diagonal elements (i.e., $g_{ij} = 0 \text{ for }
i\neq j$).  The associated scale factors ($h_i$, where $h_i^2 =
g_{ii}$) are
\begin{equation}\label{metric}
h_q = \frac{r^3}{\delta} \qquad h_p = \frac{\sin^3\theta}{\delta}
\qquad h_{\phi} = r\sin\theta
\end{equation}
With these we can immediately write down the differential elements
\begin{gather}
d\mathbf{r} = \mathbf{\hat{q}}\,\frac{r^3}{\delta}\, dq +
\mathbf{\hat{p}}\, \frac{\sin^3\theta}{\delta}\, dp +
\mathbf{\hat{\boldsymbol{\phi}}}\,r\sin\theta \,d\phi \\ d\sigma_{qp}
= \frac{r^3\sin^3\theta}{\delta^2}\,dq\,dp \qquad d\sigma_{q\phi} =
\frac{r^4\sin\theta}{\delta}\,dq\,d\phi \qquad d\sigma_{p\phi} =
\frac{r\sin^4\theta}{\delta}\,dp\,d\phi \\ d\tau =
\frac{r^4\sin^4\theta}{\delta^2}\,dq\,dp\,d\phi
\end{gather}
The Christoffel symbols of the second kind, as defined by Arfken
\cite{arfken85a}, are given by the formula
\begin{equation}
\Gamma^m{}_{ij} = \frac{1}{2}g^{km}\left(\frac{\partial g_{ik}}{\partial
x^j} + \frac{\partial g_{jk}}{\partial x^i} - \frac{\partial
g_{ij}}{\partial x^k}\right)
\end{equation}
where $g^{ii} = 1/g_{ii}$.  In matrix form they are

\begin{equation}
\Gamma^{q} = \begin{pmatrix}
-\dfrac{3r^2\cos\theta}{\delta^4}(3+5\cos^2\theta) &
\dfrac{3\sin^4\theta}{r\delta^4}(1+\cos^2\theta) & 0 \\
\dfrac{3\sin^4\theta}{r\delta^4}(1+\cos^2\theta) &
\dfrac{6\sin^6\theta\cos\theta}{r^4\delta^4}(1+\cos^2\theta) & 0 \\ 0
& 0 & \dfrac{3\sin\theta\cos\theta}{r^2}
\end{pmatrix}
\end{equation}
\begin{equation}
\Gamma^{p} = \begin{pmatrix}
-\dfrac{3r^5}{\delta^4\sin^2\theta}(1+\cos^2\theta) &
-\dfrac{6r^2\cos\theta}{\delta^4}(1+\cos^2\theta) & 0 \\
-\dfrac{6r^2\cos\theta}{\delta^4}(1+\cos^2\theta) &
-\dfrac{12\sin^2\theta\cos^2\theta}{r\delta^4}(1+\cos^2\theta) & 0 \\
0 & 0 & \dfrac{r}{\sin^2\theta}(1+\cos^2\theta)
\end{pmatrix}
\end{equation}
\begin{equation}
\Gamma^{\phi} = \begin{pmatrix} 0 & 0 &
-\dfrac{3r^2\cos\theta}{\delta^2} \\ 0 & 0 &
\dfrac{\sin^2\theta}{r\delta^2}(1-3\cos^2\theta) \\
-\dfrac{3r^2\cos\theta}{\delta^2} &
\dfrac{\sin^2\theta}{r\delta^2}(1-3\cos^2\theta) & 0
\end{pmatrix}
\end{equation}

\section{Vector Operations}

The differential operators can be derived from the metric tensor.  In
what follows $f$ is a scalar, $\mathbf{A}$ and $\mathbf{B}$ are
vectors, and $\mathsf{T}$ is a tensor.

\subsection{Gradient}
\begin{equation}
\boldsymbol{\nabla}f =
\mathbf{\hat{q}}\,\frac{\delta}{r^3}\frac{\partial f}{\partial q} +
\mathbf{\hat{p}}\,\frac{\delta}{\sin^3\theta}\frac{\partial
f}{\partial p} +
\mathbf{\hat{\boldsymbol{\phi}}}\,\frac{1}{r\sin\theta}\frac{\partial
f}{\partial \phi}
\end{equation}

\subsection{Divergence}
\begin{equation}
\boldsymbol{\nabla \cdot}\mathbf{A} =
\frac{\delta^2}{r^6}\frac{\partial}{\partial
q}\left(\frac{r^3}{\delta}A_q\right) +
\frac{\delta^2}{r^4\sin^4\theta}\frac{\partial}{\partial
p}\left(\frac{r^4\sin\theta}{\delta}A_p\right) +
\frac{1}{r\sin\theta}\frac{\partial A_{\phi}}{\partial \phi}
\end{equation}
which can also be written as either
\begin{equation}
\boldsymbol{\nabla \cdot}\mathbf{A} =
\frac{\delta^2}{r^6}\frac{\partial}{\partial
q}\left(\frac{r^3}{\delta}A_q\right) +
\frac{\delta^2}{\sin^6\theta}\frac{\partial}{\partial
p}\left(\frac{\sin^3\theta}{\delta}A_p\right) +
\frac{4}{r\delta\sin\theta}A_p + \frac{1}{r\sin\theta}\frac{\partial
A_{\phi}}{\partial \phi}
\end{equation}
or
\begin{equation}
\boldsymbol{\nabla \cdot}\mathbf{A} = \frac{\delta}{r^3}\frac{\partial
A_q}{\partial q} -\frac{3\cos\theta}{r\delta^3}(3+5\cos^2\theta)A_q+
\frac{\delta}{\sin^3\theta}\frac{\partial A_p}{\partial p} +
\frac{4}{r\delta^3\sin\theta}(1-3\cos^4\theta)A_p +
\frac{1}{r\sin\theta}\frac{\partial A_{\phi}}{\partial \phi}
\end{equation}

\subsection{Curl}
\begin{equation}
\begin{split}
\boldsymbol{\nabla \times}\mathbf{A} &=
\mathbf{\hat{q}}\,\frac{1}{r\sin\theta}\left[\frac{\delta}
{\sin^3\theta}\frac{\partial}{\partial p}(r\sin\theta A_{\phi}) 
- \frac{\partial A_p}{\partial \phi}\right]
\\ &+ \mathbf{\hat{p}}\,\frac{1}{r\sin\theta}\left[\frac{\partial
A_q}{\partial \phi} - \frac{\delta}{r^3}\frac{\partial}{\partial
q}(r\sin\theta A_{\phi})\right] \\ &+
\mathbf{\hat{\boldsymbol{\phi}}}\,\frac{\delta^2}
{r^3\sin^3\theta}\left[\frac{\partial}{\partial
q}\left(\frac{\sin^3\theta}{\delta} A_p\right) -
\frac{\partial}{\partial p}\left(\frac{r^3}{\delta}A_q\right)\right]
\\
\end{split}
\end{equation}
which is equivalent to 
\begin{equation}
\begin{split}
\boldsymbol{\nabla \times}\mathbf{A} &=
\mathbf{\hat{q}}\left[\frac{\delta}{\sin^3\theta}\frac{\partial
A_{\phi}}{\partial p} +
\frac{1-3\cos^2\theta}{r\delta\sin\theta}A_{\phi} -
\frac{1}{r\sin\theta}\frac{\partial A_p}{\partial \phi}\right] \\ &+
\mathbf{\hat{p}}\left[\frac{1}{r\sin\theta}\frac{\partial
A_q}{\partial \phi} - \frac{\delta}{r^3}\frac{\partial
A_{\phi}}{\partial q} + \frac{3\cos\theta}{r\delta}A_{\phi}\right] \\
&+
\mathbf{\hat{\boldsymbol{\phi}}}\left[\frac{\delta}{r^3}\frac{\partial
A_p}{\partial q} -\frac{6\cos\theta}{r\delta^3}(1+\cos^2\theta)A_p-
\frac{\delta}{\sin^3\theta}\frac{\partial A_q}{\partial p}
-\frac{3\sin\theta}{r\delta^3}(1+\cos^2\theta)A_q\right] \\
\end{split}
\end{equation}

\subsection{Scalar Laplacian}
\begin{equation}
\nabla^2f = \frac{\delta^2}{r^6}\frac{\partial^2f}{\partial q^2} +
\frac{\delta^2}{r^4\sin^4\theta}\frac{\partial }{\partial
p}\left(\frac{r^4}{\sin^2\theta}\frac{\partial f}{\partial p}\right) +
\frac{1}{r^2\sin^2\theta}\frac{\partial^2f}{\partial \phi^2}
\end{equation}
which can also be written as 
\begin{equation}
\nabla^2f = \frac{\delta^2}{r^6}\frac{\partial^2f}{\partial q^2} +
\frac{\delta^2}{\sin^6\theta}\frac{\partial^2f}{\partial p^2} +
\frac{4}{r\sin^4\theta}\frac{\partial f}{\partial p} +
\frac{1}{r^2\sin^2\theta}\frac{\partial^2f}{\partial \phi^2}
\end{equation}

\subsection{Vector Laplacian}
\begin{equation}
\begin{split}
\nabla^2\mathbf{A} = \mathbf{\hat{q}}\bigg[&\nabla^2A_q  + 
\frac{6\sin\theta}{r^4\delta^2}(1+\cos^2\theta)\frac{\partial
A_p}{\partial q} +
\frac{12\cos\theta}{r\delta^2\sin^3\theta}(1+\cos^2\theta)\frac{\partial
A_p}{\partial p} + \frac{6\cos\theta}{r^2\delta\sin\theta}\frac{\partial
A_{\phi}}{\partial \phi} 
\\
&-\frac{9}{r^2\delta^4}(1+3\cos^2\theta+4\cos^4\theta)A_q
-\frac{12\cos\theta}{r^2\delta^4\sin\theta}(1+3\cos^4\theta)A_p\bigg] \\
+\mathbf{\hat{p}}\bigg[&\nabla^2A_p -
\frac{6\sin\theta}{r^4\delta^2}(1+\cos^2\theta)\frac{\partial
A_q}{\partial q} -
\frac{12\cos\theta}{r\delta^2\sin^3\theta}(1+\cos^2\theta)\frac{\partial
A_q}{\partial p} -
\frac{2}{r^2\delta\sin^2\theta}(1-3\cos^2\theta)\frac{\partial
A_{\phi}}{\partial \phi} \\ 
&+\frac{18\sin\theta\cos\theta}{r^2\delta^4}(1+\cos^2\theta)A_q
-\frac{2}{r^2\delta^4\sin^2\theta}(5+3\cos^2\theta-9\sin^2\theta\cos^4\theta)A_p \bigg]\\
+\mathbf{\hat{\boldsymbol{\phi}}}\bigg[&\nabla^2A_{\phi} - \frac{6\cos\theta}{r^2\delta\sin\theta}\frac{\partial A_q}{\partial \phi} + \frac{2}{r^2\delta\sin^2\theta}(1-3\cos^2\theta)\frac{\partial A_p}{\partial \phi} - \frac{A_{\phi}}{r^2\sin^2\theta}\bigg]
\end{split}
\end{equation}

\subsection{Directional Derivative}
\begin{equation}
\begin{split}
(\mathbf{A}\boldsymbol{\cdot\nabla})\mathbf{B} =
\mathbf{\hat{q}}\bigg[&\mathbf{A}\boldsymbol{\cdot\nabla}B_q +
\frac{3\sin\theta}{r\delta^3}(1+\cos^2\theta)A_qB_p +
\frac{6\cos\theta}{r\delta^3}(1+\cos^2\theta)A_pB_p +
\frac{3\cos\theta}{r\delta}A_{\phi}B_{\phi}\bigg]\\
+\mathbf{\hat{p}}\bigg[&\mathbf{A}\boldsymbol{\cdot\nabla}B_p -
\frac{3\sin\theta}{r\delta^3}(1+\cos^2\theta)A_qB_q -
\frac{6\cos\theta}{r\delta^3}(1+\cos^2\theta)A_pB_q \\
&-\frac{1}{r\delta\sin\theta}(1-3\cos^2\theta)A_{\phi}B_{\phi}\bigg]\\
+\mathbf{\hat{\boldsymbol{\phi}}}\bigg[&\mathbf{A}
\boldsymbol{\cdot\nabla}B_{\phi}
- \frac{3\cos\theta}{r\delta}A_{\phi}B_q +
\frac{1}{r\delta\sin\theta}(1-3\cos^2\theta)A_{\phi}B_p\bigg]
\end{split}
\end{equation}

\subsection{Divergence of a Tensor}
\begin{equation}
\begin{split}
\boldsymbol{\nabla\cdot}\mathsf{T} =
\mathbf{\hat{q}}\bigg[&\boldsymbol{\nabla\cdot}
(\mathsf{T}_{qq}\mathbf{\hat{q}}
+\mathsf{T}_{pq}\mathbf{\hat{p}} + \mathsf{T}_{\phi
q}\mathbf{\hat{\boldsymbol{\phi}}}) +
\frac{3\sin\theta}{r\delta^3}(1+\cos^2\theta)\mathsf{T}_{qp} +
\frac{6\cos\theta}{r\delta^3}(1+\cos^2\theta)\mathsf{T}_{pp} +
\frac{3\cos\theta}{r\delta}\mathsf{T}_{\phi \phi}\bigg] \\
+\mathbf{\hat{p}}\bigg[&\boldsymbol{\nabla\cdot}
(\mathsf{T}_{qp}\mathbf{\hat{q}} +
\mathsf{T}_{pp}\mathbf{\hat{p}} + \mathsf{T}_{\phi p}
\mathbf{\hat{\boldsymbol{\phi}}}) -
\frac{3\sin\theta}{r\delta^3}(1+\cos^2\theta)\mathsf{T}_{qq} -
\frac{6\cos\theta}{r\delta^3}(1+\cos^2\theta)\mathsf{T}_{pq} \\
&-\frac{1}{r\delta\sin\theta}(1-3\cos^2\theta)\mathsf{T}_{\phi \phi}
\bigg]\\
+\mathbf{\hat{\boldsymbol{\phi}}}\bigg[&\boldsymbol{\nabla\cdot}
(\mathsf{T}_{q\phi}\mathbf{\hat{q}} +
\mathsf{T}_{p\phi}\mathbf{\hat{p}} + \mathsf{T}_{\phi
\phi}\mathbf{\hat{\boldsymbol{\phi}}}) - 
\frac{3\cos\theta}{r\delta}\mathsf{T}_{\phi q} +
\frac{1}{r\delta\sin\theta}(1-3\cos^2\theta)\mathsf{T}_{\phi p}\bigg]
\end{split}
\end{equation}

\section*{Acknowledgments}
This work was supported by the Office of Naval Research.


\end{document}